\documentclass[a4paper,11pt]{article}

\usepackage{cite}
\usepackage{authblk}

\usepackage{amsmath}
\usepackage{amscd}
\usepackage{amsfonts}
\usepackage{amssymb}

\begin{document}
\title{\bf{Exact results in two dimensional chiral hydrodynamics with gravitational anomalies\vspace{1em}}}
\author{
{\bf {\normalsize Rabin Banerjee}}\\
 {\normalsize \it S.~N.~Bose National Centre for Basic Sciences,}\\
 {\normalsize \it Block-JD, Sector III, Salt Lake, Kolkata-700098, India.}\\ 
 {\texttt{rabin@bose.res.in}}}

\date{}

\maketitle

\vspace{-1em}
\begin{abstract}
An exact formulation of two dimensional chiral hydrodynamics with diffeomorphism and conformal anomalies is provided. The constitutive relation involving the stress tensor is computed. It reveals a one parameter class of solutions which is a new result. For a particular value of this parameter, the results found in the gradient expansion scheme are reproduced. Moreover, the constitutive relation is analogous to the corresponding relation for an ideal fluid, appropriately modified to include the chirality property, which has also been derived here. 
\end{abstract}
\section{Introduction}

A universal effective picture of any finite temperature quantum field theory on large time and length scales is provided by hydrodynamics\cite{liflan}. This has led to a revival of interest in this topic. Also, the fluid-gravity map of AdS/CFT has further bolstered this activity. The fundamental equations governing the dynamics in this description are the conservation laws manifesting the global symmetries of the underlying theory. While the space-time symmetry leads to the conservation of the stress tensor, the charge  symmetry leads to the conservation of charge current. Furthermore, the conservation laws are supplemented by constitutive relations expressing the stress tensor and the current in terms of the basic fluid variables which are velocity, temperature and chemical potential. These constitutive relations, naturally, have to be consistent with the conservation laws and are thus usually obtained by requiring compatibility with the second law of thermodynamics\cite{liflan}. This standard picture is nontrivially modified in the presence of quantum anomalies.

 In this paper we investigate the structure of constitutive relations in the presence of gravitational anomalies in two dimensions. Using  techniques peculiar to two dimensions we obtain new results with novel interpretations and implications. We consider a chiral theory having both diffeomorphism and trace anomalies. The constitutive relation which is exact and not any order by order expansion, accounts for both these anomalies. Incidentally, two dimensions are very special and merit a separate study that is distinct from an arbitrary dimensional analysis. The metric in two dimensions is conformally flat which ensures that the effective action is exactly obtainable. A derivative expansion approach that is appropriate for calculating the effective action in higher dimensional fluid dynamics\cite{arxiv:1203,jain,superfluid} now becomes redundant. In fact much of the two dimensional physics is hidden if one adopts the standard derivative expansion method or its variations\cite{wesszumino, valle, cones,oz}.
 
 In section $2~$ we provide a brief overview of chiral gravitational anomalies in two dimensions\cite{new} highlighting the role of the diffeomorphism anomaly vis-a-vis the conformal (trace) anomaly. This is necessary to appreciate the distinction between a chiral anomalous theory and a nonchiral anomalous theory. Section $3~$contains the constitutive relation involving the stress tensor. An exact relation is obtained that is put in the form of the corresponding relation for  an ideal chiral fluid. It may be observed that the constitutive relation for an ideal chiral fluid is different from a normal one. This is explained and the relevant relation is derived in section $4$. The last section contains our concluding remarks. 
\section{General notions on gravitational anomalies}

Consider a relativistic quantum field theory on a manifold with a time like killing vector. Since the manifold has a time like killing vector, the metric can be chosen to be static. By an appropriate coordinate redefinition, the metric on such a manifold can be expressed as, 
\begin{equation}
\label{met}
 ds^2=-e^{2\sigma(\bar x)}(dt~+~a_{i}(\bar x)~dx^{i} )^{2}~ + g_{ij}dx^{i}dx^{j}  
\end{equation}
where the coordinates $~\bar x = x^{i}~(i=1,2,3...,p) ~$ specify $(p)$ spatial degrees of freedom. For the special case of (1+1) dimensions, the Riemann manifold is conformally flat and we may write 
\begin{equation}
\label{metmod}
 ds^2~=~-e^{2\sigma(r)}dt^{2}~ + g_{11}(r)dr^{2}  
\end{equation}
where the above form is taken for its resemblance with (\ref{met}) and also for  comparing with results given earlier.
 Precisely because two dimensional Riemann spaces may be put in the form (\ref{metmod}), it is feasible to explicitly calculate the effective action. Variation of the effective action  immediately yields the exact expression for the energy momentum tensor. This should be contrasted with higher dimensions where the effective action and hence, the stress tensor, is obtained approximately as a power series expansion. Nevertheless even though the stress tensor may not have a closed form, its conservation  law (whether normal or anomalous) is obtained as an exact result. 
 
   This naturally leads us to the topic of anomalies. An anomaly is a breakdown of classical symmetry upon quantization. Generally this is manifested as a clash of  symmetries. For instance, in QED if a gauge invariant regularisation is chosen, then axial symmetry is violated leading to non-conservation of the axial current. Likewise, it is  possible to preserve conservation of axial current by using a different regularisation  but then gauge invariance gets violated. The essence of the anomaly is that there is no regularisation that simultaneously preserves both the vector and axial symmetries.
   
   A similar feature holds for the gravitational case. Here there is a clash between general coordinate invariance and conformal invariance. Preserving the first yields covariant conservation of the stress tensor  whose trace, however, is non zero. A vanishing  trace follows by involving a conformal invariant regularisation  but this leads to a non conservation of the stress tensor. Since  general coordinate invariance is regarded as more fundamental, this is usually kept intact at the expense of a trace anomaly. The above comments which are valid for a vector theory, have to be modified if the theory is chiral(i.e. chiral fermions moving in a curved background ). In this case both general coordinate invariance  and conformal invariance are violated. There is a diffeomorphism anomaly (non conservation of the stress tensor) as well as a trace anomaly. This point is usually not appreciated leading to confusion. Moreover, for a chiral theory, there is a further subtlety. Chiral expressions may be regularised in various ways but there are particularly two which are useful for different reasons. Using a covariant regularisation one may obtain covariant expressions. However these do not satisfy  the so called Wess-Zumino consistency condition  or the integrability condition and hence cannot be directly obtained from a variation of the effective action. Expressions that  are derived  in this manner are termed `consistent'. They satisfy the Wess-Zumino  consistency  condition  but do not transform covariantly under a general coordinate  transformation.\footnote{Covariant and consistent forms are related by local counterterms\cite{zumino, banerjee}} For gravitational theories covariant expressions  are preferred otherwise the basic transformation laws of tensors are violated.
   
   Since we will be dealing with a chiral theory, let us  analyse it in more details. For chiral fermions in a  two dimensional  gravitational  background the possible structures of the covariant diffeomorphesm anomaly and the trace anomaly are uniquely  fixed, upto some normalisation, from general considerations as,
   \begin{equation}
   \label{trcovener}
   \triangledown_{\mu}T^{\mu}_\nu  = N_{d} \bar\epsilon_{\nu\mu}\nabla^{\mu}R,~~ T_\mu ^{\mu} ~=~ N_{t}R
   \end{equation}
 where $~ N_{d}~$and$~  N_{t} ~$ are normalization constants for the diffeomorphism  anomaly and trace anomaly, respectively. Here R is the Ricci scalar which for the metric(\ref{metmod}) is given by, 
    \begin{equation}
    \label{ricci}
 R= \dfrac{1}{g_{11}^2}(g_{11}'\sigma ' - 2g_{11} \sigma^{'2} - 2g_{11} \sigma^{'' })
 \end{equation}
and $~\bar\epsilon_{\mu\nu}=\sqrt{-g}\epsilon_{\mu\nu}~$ , with $g=\text{det}\,g_{\mu\nu}$ and  $~\epsilon_{\mu\nu}~~$ is the antisymmetric numerical tensor    $~\epsilon_{01}=-\epsilon_{10}=-1$. The functional form for the anomalies given in (\ref{trcovener}) is dictated by dimensionality, covariance and parity. The stress tensor $T_{\mu\nu}$, due to its chiral nature, satisfies the property \cite{kulkarni}.
 \begin{equation}
 \label{enmom}
 T _{\mu\nu} = -\dfrac{1}{2}(\bar\epsilon_{\mu\rho}T _{\nu}^{\rho}~+~\bar\epsilon_{\nu\rho}T _{\mu}^{\rho}) ~ +~ \dfrac{1}{2}g_{\mu\nu}T _{\alpha}^{\alpha}
 \end{equation}
 which is a generalisation of the property satisfied, for example, by a chiral current ,
 \begin{equation}
 \label{chicur}
 J_{\mu}~=~-\bar\epsilon_{\mu\nu}J^{\nu}
 \end{equation}
The chiral property is best manifested in the null coordinates, 
\begin{equation}
\label{null}
u = t- r^{*}  ~,~  v = t + r^{*}~,~\dfrac{dr}{dr^{*}}~=~-\dfrac{e^{\sigma}}{\sqrt{g_{11}}}
\end{equation}
In these coordinates the metric is given by, 
\begin{equation}
 \label{metnull}
ds^{2}= -\dfrac{1}{2} e^{2\sigma} (dudv~+~dvdu)
\end{equation}
and the antisymmetric tensor$~\bar\epsilon_{\mu\nu}$ is given by 
\begin{equation}
\label{barepsiuv}
\bar\epsilon_{uv}~= \dfrac{e^{2\sigma}}{2}
\end{equation}
In our chosen system of coordinates it is easy to check from (\ref{enmom}) that chirality is manifested by the vanishing of $vv$-component while, expectedly, the $uv$-component yields the trace
\begin{equation}
\label{eneruv}
T_{vv}~=~0,~~ T_{uv}= \dfrac{g_{uv}}{2}T^{\alpha}_{\alpha}=-\dfrac{e^{2\sigma}}{4}T^{\alpha}_{\alpha}
\end{equation} 
It is interesting to observe that the above results follow from general  considerations. Subsequently we shall reproduce these from a direct calculation of the stress tensor obtained from the chiral effective action.

 We now reveal a connection between the normalization constants$~ N_{d}~$and$~ N_{t} ~ $ appearing in (\ref{trcovener}). For $\nu~ =~v $, the left hand side of the first relation in(\ref{trcovener}) becomes,
 \begin{equation}
 \label{coveneruv}
 \nabla_{\mu}T^{\mu}_{v} = \nabla_{u}T^{u}_{v} ~+ \nabla_{v}T^{v}_{v} = \nabla_{u}(g^{uv} T_{vv})~+~\nabla_{v}(g^{vu} T_{uv})
 = \nabla_{v}[(-\dfrac{2}{e^{2\sigma}})(-\dfrac{e^{2\sigma}}{4})T^{\alpha}_{\alpha}]
 \end{equation} 
 where the results (\ref{eneruv} ) have been exploited. Substituting the expression for the trace$~T _{\alpha}^{\alpha}~$ from(\ref{trcovener})we find,
 \begin{equation}
 \label{coveneruv1}
 \nabla_{\mu}T^{\mu}_{v} =\dfrac {N_{t}}{2} \nabla_{v}R
 \end{equation}
 Now the right hand side of the first relation in (\ref{trcovener}) for $\nu= v$, simplifies to,
 \begin{equation}
 \label{coveneruv2}
\nabla_{\mu}T^{\mu}_{v} =N_{d}\bar\epsilon_{v\mu}\nabla^{\mu}R= - N_{d}\dfrac{e^{2\sigma}}{2}\nabla^{u}R=N_{d}\nabla_{v}R
 \end{equation}
 Equating (\ref{coveneruv1}) and (\ref{coveneruv2}) immediately yields the cherished connection,
 \begin{equation}
 \label{trdifnor}
 N_{t}=2N_{d}
\end{equation} 
 It is now important to stress that chirality enforces both the trace and diffeomorphism anomalies. Note that the trivial(anomaly free) case$~N_{t}=N_{d}=0$ is ruled out since, using the unidirectional property of chirality, it is feasible to prove the existence of the diffeomorphism anomaly in$~1+1~$ dimensions\cite{fulling}.This  implies that both $N_{d}$ and $N_{t}$ are non zero, connected by the relation (\ref{trdifnor}). For  a non chiral theory, on the contrary,  the relation (\ref{trdifnor}) does not hold and the absence of a unidirectional property leads to a scenario where there is no diffeomorphism anomaly but a trace anomaly exists or vice-versa.
 This is the crucial distinction between a chiral and a non chiral theory. In any formulation of the chiral theory, therefore , it is imperative to properly account for both diffeomorphism  and trace anomalies
    - a feature that we claim  has not been accounted for or emphasized in the literature on hydrodynamics with anomalies.
    
     The general analysis is now supplemented by explicit structures. The two dimensional chiral effective action is defined as,\cite{leut,2008}
    \begin{equation}
    \label{ac}
     Z(\omega)~=~ -\dfrac{1}{12\pi} \int d^{2}xd^{2}y \epsilon^{\mu\nu}\partial_{\mu}\omega_{\nu}(x)\triangle^{-1}(x,y)\partial_{\rho}[(\epsilon^{\rho\sigma}+ \sqrt{-g}~g^{\rho\sigma})\omega_{\sigma}(y)]
     \end{equation}
     where $~\triangle^{-1}~$ is the inverse of the  d'alembertian $~\triangle~=~ (\nabla^{\mu}\nabla_{\mu})
     ~=~\dfrac{1}{\sqrt -g}\partial_{\mu}[{\sqrt -g}~ g^{\mu\nu}\partial_{\nu}]~$ and $~\omega_{\mu}~$ is the spin connection. This is an exact result. It is basically the determinant of the Weyl operator $D=\sigma^a {e}^{\mu}_a (\partial_\mu + i\omega_\mu)$ where the expression has been appropriately normalised to include the chiral(half) factor.
     
     The stress tensor may now be obtained by a variation of this effective action. That would yield the consistent expression. The covariant form that is relevant  for our purpose is obtained by adding a local polynomial. This is allowed since energy momentum tensors and currents are only defined  modulo local polynomials which manifest the regularisation freedom. We obtain,
     \begin{equation}
     \label{varac}
     \delta Z ~=~ \int d^{2}x \sqrt {-g}~(\dfrac{1}{2}\delta g_{\mu\nu} T^{\mu\nu})+ l
     \end{equation}
     where  the local polynomial is given by,
     \begin{equation}
     \label{poly}
     l = -\dfrac{1}{12\pi}\int d^{2}x \epsilon^{\mu\nu} [\omega_{\mu}\delta\omega_{\nu} + \dfrac{1}{8}R e^{a}_{\mu}\delta e^{a}_{\nu }]
\end{equation}      
where the zweibein vectors $~e ^{a}_{\mu}~$ fix the metric $~ g_{\mu\nu}~$.

The covariant energy momentum tensor is easily read off  from the above relations as,
\begin{equation}
\label{enermom}
T^{\mu}_{\nu} ~=~ \dfrac{1}{96\pi} (\dfrac{1}{2}D^{\mu}G D_{\nu}G - D^{\mu}D_{\nu}G + \delta^{\mu}_{\nu}R)
\end{equation}
The chiral nature is highlighted by the presence of the chiral covariant  derivative $~D_{\mu}~$ defined in terms of the ordinary  covariant derivative $~\nabla_{\mu}~$,
\begin{equation}
\label{chicovder}
D_{\mu}= ~\nabla_{\mu}~ - \bar\epsilon_{\mu\nu}~\nabla^{\nu}~=- \bar\epsilon_{\mu\nu}~ D^{\nu}
\end{equation}
while R is defined in (\ref{ricci}). The auxiliary field G  in (\ref{enermom}) is given by ,
\begin{subequations}
\begin{align}
\label{auxg}
G(x)~=~ \int d^{2}y\sqrt{-g}~ \triangle ^{-1}(x,y)R(y)
\end{align}
so that
\begin{align}
\label{alemg}
\triangle G(x) = R(x)
\end{align}
\end{subequations}
Calculating  the covariant divergence and the trace of $~T ^{\mu}_{\nu}~$ from (\ref{enermom}) we find,
\begin{equation}
\label{trenerr}
\nabla_{\mu} T^{\mu}_{\nu}=~\dfrac{\bar\epsilon_{\nu\mu}}{96\pi}\nabla^{\mu}R, ~~T^{\mu}_{\mu}=\dfrac{R}{48\pi}
\end{equation}
These are the covariant expressions for the diffeomorphism anomaly and the trace anomaly\cite{new}.They agree with the structures (\ref{trcovener},\ref{trdifnor}) given on general grounds.Furthermore, exploiting (\ref{chicovder}) it is  simple to establish the chirality property (\ref{enmom}) following from (\ref{enermom}).

 Solving (\ref{alemg}) we obtain,
\begin{equation}
\label{solg}
 G= G_{0}(r) -4pt +q,~ ~~\partial_{r} G_{0}~=~\dfrac{\sqrt g_{11}}{e^ \sigma}(-\dfrac{2\sigma{'}e^{\sigma}}{\sqrt g_{11}} + z)
 \end{equation}
  where $p~$,$q$ and $z$ are constants. 

As a consistency check we show how the general solution (\ref{solg}) yields the results for flat space which may be directly obtained from (\ref{alemg}). The covariant d'alembertian $\triangle$ simplifies to $\partial^{2}$ while the Ricci scalar $R$ vanishes. Then (\ref{alemg}) is just the massless Klein-Gordon equation,
\begin{equation}
\label{23}
\partial^{2}G=0
\end{equation}
In the null variables this equation takes the form, 
\begin{equation}
\label{24}
\partial_{u}\partial_{v}G(u,v)=0
\end{equation}
whose general solution is expressed as a sum of the left($L$) and right($R$) moving modes,
\begin{equation}
\label{25}
G(u,v)= \phi_{L}(u)+\phi_{R}(v)
\end{equation}
To compare this solution with (\ref{solg}) we rewrite it in flat space using the null coordinates (\ref{null}),
\begin{equation}
\label{26}
G=-\dfrac{z}{2}(v-u)-\dfrac{4p}{2}(u+v)+Q
 = (\dfrac{z}{2}-2p)u-(\dfrac{z}{2}+2p)v+Q 
\end{equation}
and Q is a constant given by,
 \begin{equation}
\label{27}
Q=q+zk+k_{0}
\end{equation}
The constants q and z occur in (\ref{solg}) while the other constants $k_{0}, k$ are the integration constants related to the solutions of $G_{0}$ and $r$, 
\begin{equation}
\label{28}
G_{0}=\int zdr=zr+k_{0}, r=-\int dr^{*}=-r^{*}+k
\end{equation}
Comparing (\ref{25}) with (\ref{26}) yields the desired identification,
\begin{equation}
\phi_{L}(u)=(\dfrac{z}{2}-2p)u+\dfrac{Q}{2}, \\
\phi_{R}(v)=-(\dfrac{z}{2}+2p)v+\dfrac{Q}{2}
\end{equation}
  It is now possible to obtain the  explicit form for $~T^{\mu}_{\nu}~$. In the null coordinates we have already found (see (\ref{eneruv})) that $~T_{vv}~=0~$ and $~T_{uv}~$ is proportional to the trace. Using (\ref{eneruv}) and (\ref{trenerr}) we get, 
  \begin{equation}
  \label{eneruv1}
  T_{uv}= -\dfrac{e^{2\sigma}}{4}\times \dfrac{R}{48\pi} = -\dfrac{e^{2\sigma}R}{192\pi}
  \end{equation}
   The only other component $~T_{uu}~$ is obtained from (\ref{enermom}) and(\ref{solg}) after a slight calculation,
   \begin{equation}
   \label{eneruv2}
   T_{uu}=\dfrac{1}{96\pi}\dfrac{e^{2\sigma}}{g_{11}^2}(2g_{11}\sigma{''}-\sigma{'}g_{11}{'})+C
   \end{equation}
 where $C$ is a constant defined as,
\begin{eqnarray}
\label{constant}
&& C= \frac{1}{12\pi}\left(p-\frac{z}{4}\right)^2
\end{eqnarray} 
 Observe that although there are three distinct constants $p$, $q$ and $z$, the physically relevant constant is a combination of these\footnote{The constant $q$, absent in (\ref{constant}), occurs in the construction of the gauge current which is not considered here.}. That only one constant appears is a consequence of the fact that the stress tensor $T^\mu_\nu$ is a solution of the anomaly equation (\ref{trenerr}) which is a linear differential equation. This constant may be fixed by choosing a boundary condition. However to keep the discussion general, we retain $C$. Relations (\ref{eneruv1}),(\ref{eneruv2}) along with $T_{vv}~=0~$ yield  the exact expressions for the covariant stress tensor in the null coordinates.
\section{Anomaly induced constitutive relation}

The results for the stress tensor will now be used to deduce the constitutive relation. This relation is obtained by expressing the stress tensor in terms of the fluid variables. It is conveniently done by adopting the comoving frame. In this frame the fluid velocity  $ u^{\mu} $ is normalised as,
\begin{equation}
\label{A}
u^{\mu}u_{\mu}=-1.
\end{equation}
Note that the norm of the velocity has to be negative since we are considering time like trajectories. The absolute nomalisation is fixed to unity by taking the comoving frame. Subjected to the condition (\ref{A}) and the metric (\ref{metmod}), the usual ansatz for the velocity follows,
 \begin{equation}
 \label{comovvel}
 u^{\mu}= e^{-\sigma(r)}(1,0),~ u_{\mu}=~ -e^{\sigma(r)}(1,0), ~~(\mu = t,r ).
 \end{equation}
 Since the metric is static, expectedly the above ansatz for the velocity is also time independent. Correspondingly, in the null coordinates, we have,
 \begin{equation}
 \label{comovuv}
 u_{\mu}= -\dfrac{e^{\sigma(r)}}{2}(1,1),~~ u^{\mu}=~ e^{-\sigma(r)}(1,1),~~ (\mu =u,v)
\end{equation}  
 which has a more symmetrical structure. Introduce the dual vectors,
 \begin{equation}
 \label{comovdual}
 \tilde{u_{\mu}}= \bar\epsilon_{\mu\nu}u^{\nu}=\dfrac{e^{\sigma(r)}}{2}(1,-1),
 ~~\tilde{u^{\mu}}= \bar\epsilon^{\mu\nu}u_{\nu}= e^{-\sigma(r)}(1,-1),~~ (\mu =u,v)
 \end{equation}
where we have used (\ref{barepsiuv}) to obtain the explicit values.Observe that the dual velocity fields satisfy a different normalization from the original ones, since $~\tilde{u^{\mu}}\tilde{u_{\mu}}=1~$.
 
 To obtain the constitutive relation let us first write certain relations that express covariant  combinations of the fluid variables in terms of the metric coefficients,
 \begin{equation}
 \label{ugrad}
 u^{\mu}\nabla^{\rho}\nabla_{\mu}u_{\rho}= -\dfrac{1}{2}R,
 ~~u^{\rho}\nabla^{\mu}\nabla_{\mu}u_{\rho}~=~\dfrac{\sigma^{'2}}{g_{11}}
 \end{equation}
 Using the above equalities, the stress tensor (\ref{eneruv1}, \ref{eneruv2}) in the null coordinates is expressed as
 \begin{equation}
 \label{tuu}
 T_{uu}=\dfrac{e^{2\sigma}}{48\pi}( u^{\mu}\nabla^{\rho}\nabla_{\mu}u_{\rho}-u^{\rho}\nabla^{\mu}\nabla_{\mu}u_{\rho})+C,
 ~~T_{uv}= \dfrac{e^{2\sigma}}{96\pi}( u^{\mu}\nabla^{\rho}\nabla_{\mu}u_{\rho})
 \end{equation}
These relations along with $~T_{vv}=0~$, are expressible as follows:
\begin{eqnarray}
\label{tmunu}
T_{\mu\nu}= \left\lbrace\dfrac{1}{48\pi}[(u^{\alpha}\nabla^{\beta }-u^{\beta}\nabla^{\alpha })\nabla_{\alpha}u_{\beta}]+e^{-2\sigma}C\right\rbrace (2u_{\mu}u_{\nu}-\tilde{u_{\mu}}u_{\nu}~-~u_{\mu}\tilde{u_{\nu}})\nonumber\\
-\left\lbrace\dfrac{1}{48\pi}(u^{\beta}\nabla^{\alpha }\nabla_{\alpha}u_{\beta})-e^{-2\sigma}C\right\rbrace g_{\mu\nu}
\end{eqnarray}

  The tensorial character of this and other subsequent relations implies their general validity irrespective of the choice of a particular ansatz like (\ref{comovvel}). This is also clear from the observation that the expressions for the gravitational and trace anomalies are correctly reproduced, a point that is elaborated at the end of this section. Of course wherever the variable $\sigma(r)$ appears explicitly, as in the term $e^{-2\sigma}C$, it has to be taken as time independent which just follows from the choice of our original metric (\ref{metmod}).

Now hydrodynamics has to be formulated in terms of the local fluid velocity and the local temperature. One would thus expect the appearance of two independent local degrees of freedom. However the fact that there is a further constraint (\ref{A}) implies the existence of only one such degree of freedom. Moreover, since the velocity is already a function of $ \sigma $ (see (\ref{comovvel})), the temperature should also be expressed through the same function. The only way this can be done in (\ref{tmunu}) is to capitalise on the term $e^{-2\sigma} C $.  Introducing the locally (red shifted) temperature T by the usual Tolman relation, $T=T_{0}e^{-\sigma}$, we obtain,
\begin{eqnarray}
\label{CT}
T_{\mu\nu}= \left\lbrace\dfrac{1}{48\pi}[(u^{\alpha}\nabla^{\beta }-u^{\beta}\nabla^{\alpha })\nabla_{\alpha}u_{\beta}]+\bar C T^{2}\right\rbrace (2u_{\mu}u_{\nu}-\tilde{u_{\mu}}u_{\nu}-u_{\mu}\tilde{u_{\nu}})-\nonumber\\~\left\lbrace\dfrac{1}{48\pi}(u^{\beta}\nabla^{\alpha }\nabla_{\alpha}u_{\beta})-\bar C T^{2}\right\rbrace g_{\mu\nu}
\end{eqnarray}
where $ \bar C = CT_{0}^{-2}$ is an arbitrary constant.
 This is the cherished constitutive relation in the presence of chiral(covariant) anomalies. It is an exact relation and is a new result. There are no derivative corrections. Moreover, the specific functional form of the factor involving the velocity is  unique since it is the only one that ensures chirality. This will be further clarified in the next section. The constant C in (\ref{tmunu}) is rewritten in terms of another constant $\bar{C}$ \footnote{Note that $T_0$ is also a constant. Here it is the equilibrium temperature. In black hole physics it is the Hawking temperature.} to facilitate comparison with existing results in the literature, as explained later. A relation similar to (\ref{CT}) was obtained in\cite{cones}, where the coefficient $\bar{C}$ was fixed by relating it to the coefficient of the trace anomaly.

  It may be observed that the appearance of the fluid temperature in the constitutive relation (\ref{CT}) is essentially linked with the presence of the integration constant. This also explains the apparent lack of any dependence of $T_{0}$ in the effective action (\ref{ac}). The point is that it is actually hidden in the nonlocal structure of (\ref{ac}). In passing over to local expressions it is mandatory to introduce the additional fields $G$ which satisfy a differential equation. The solution of $G$ thus involves an integration constant which is eventually instrumental in introducing the fluid temperature.
 
  In the fluid variables  it is straightforward to show that (\ref{tmunu})or (\ref{CT}) correctly reproduces both the diffeomorphism anomaly (\ref{trenerr}) as well as the trace anomaly (\ref{trenerr}) obtained earlier from the equilibrium partition function. For example the trace anomaly is trivially seen by contracting(\ref{tmunu}) and using $~u^{\mu}u_{\mu}=-1~$, $~\tilde{u^{\mu}}u_{\mu}=0~$to yield,
 \begin{equation}
 \label{tmumu}
 T^{\mu}_{\mu}= -\dfrac{1}{24\pi}(u^{\alpha}\nabla^{\beta }\nabla_{\alpha}u_{\beta})
 \end{equation}
  This reproduces, on using (\ref{ugrad}), the  expression (\ref{trenerr}) for the trace anomaly. Likewise the diffeomorphism anomaly may also be reproduced.
\section{Correspondence with ideal chiral fluid constitutive relation}
  
  We now show that the relation (\ref{tmumu}) can be put exactly in the form governing an ideal chiral  fluid. Let us recall that the familiar(textbook) ideal fluid  constitutive relation is given by the well known formula 
  \begin{equation}
  \label{consti}
  T^{0}_{\mu\nu}~=~ (\epsilon +p)u_{\mu}u_{\nu} + pg_{\mu\nu}
\end{equation}    
where $~\epsilon~$ and $p ~$are the energy density and pressure, respectively. 

Just as the above relation(which is the zeroth order contributing in a gradient expansion) cannot account for dissipative effect it also fails to account for chirality. Here we consider the modifications brought about by chirality. In two dimensions chiral vectors or tensors must satisfy properties similar to (6) or (5). Note that chirality is an algebraic property which should be manifested at the classical level itself. To incorporate these features in (\ref{consti}), replace the velocity $~u_{\mu}~$ by the chiral velocity vector,
 \begin{equation}
 \label{chiu}
 u^{c}_{\mu}= ~u_{\mu}-~\tilde{u_{\mu}}~=~u_{\mu}~ -\bar\epsilon_{\mu\nu}u^{\nu}
\end{equation}    
  It is easy to see that (\ref{chiu}) satisfies the chiral condition 
  \begin{equation}
  \label{chiu1}
  u^{c}_{\mu}= -\bar\epsilon_{\mu\nu}u^{\nu(c)}
  \end{equation}
  which is the analogue of (6). Then the constitutive relation (\ref{consti}) is replaced by the form,
    \begin{equation}
    \label{chit}
  T_{\mu\nu}~=~ (\epsilon^{c} +p^{c})u_{\mu}^{c}u_{\nu}^{c} + p^{c}g_{\mu\nu}
\end{equation} 
Where $\epsilon^{c}$ and $p^{c}$ may be interpreted as the analogous of $\epsilon$ and $p$ in (\ref{consti}) for anomalous(chiral) sector. This stress tensor  satisfies the chirality condition
   \begin{equation}
   \label{chit1}
 T _{\mu\nu} = -\dfrac{1}{2}(\bar\epsilon_{\mu\rho}T _{\nu}^{\rho}~+~\bar\epsilon_{\nu\rho}T _{\mu}^{\rho}) ~ +~ p^{c}{g_{\mu\nu}}.
 \end{equation}
 Using the property, 
 \begin{equation}
 \label{uu}
 u_{\mu}^{c}u^{\mu c}~=~0
 \end{equation}
 following from the basic definition (\ref{chiu}) or (\ref{chiu1}), it follows from (\ref{chit}) that $p^{c}$ gets identified with the trace of the stress tensor,
 \begin{equation}
 \label{pr}
 p^{c}~=~ \dfrac{1}{2} (T^{\mu}_{\mu})
 \end{equation}
 Putting this back in (\ref{chit1}) immediately yields the standard chiral property (\ref{enmom}) for $~T_{\mu\nu}~$.
 
  The total stress tensor is a sum of the usual(non-anomalous) part (\ref{consti})and the chiral(anomalous) part(\ref{chit1}), $T^{total}_{\mu \nu}=T^{0}_{\mu \nu}+T_{\mu \nu}$. The diffeomorphism invarient part of the total effective action yields $T^{0}_{\mu \nu}$ (satisfying $\triangledown^{\mu} T_{\mu \nu}=0 $) while the anomalous part yields $T_{\mu \nu}$ (satisfying (\ref{trenerr})).
  
  It is now possible to recast (\ref{CT}) precisely in the form (\ref{chit}) with appropriate identifications for $\epsilon^{c}$ and $p^{c}$. Using (\ref{chiu}) the relation (\ref{chit}) is  expressed  as
   \begin{equation}
   \label{mufin}
  T_{\mu\nu}~=~ (\epsilon +p)(2u_{\mu}u_{\nu} - u_{\mu}\tilde{u_{\nu}} -\tilde{u_{\mu}}u_{\nu})~+~(\epsilon +2p)g_{\mu\nu}
\end{equation} 
where, at an intermediate step, the identity, 
\begin{equation}
\label{tiumu}
\tilde{u_{\mu}}\tilde{u_{\nu}} - u_{\mu}u_{\nu}~=~ g_{\mu\nu}
\end{equation}
has been exploited. We regard (\ref{mufin}) as the appropriate  constitutive  relation for an ideal chiral fluid in two dimensions. 

As announced earlier the constitutive relation (\ref{tmunu}) in the presence of anomalies has precisely the same form (\ref{mufin}) as the ideal case once we identify, $$\epsilon^{c} +p^{c} = \dfrac{1}{48\pi}(u^{\alpha}\nabla^{\beta}~-~u^{\beta}\nabla^{\alpha})\nabla_{\alpha}u_{\beta}+ \bar{C}T^{2}$$
\begin{equation}
\label{ep}
\epsilon^{c} +2p^{c} = -\dfrac{1}{48\pi}~u^{\beta}\nabla^{\alpha}\nabla_{\alpha}u_{\beta}+\bar{C}T^{2}
\end{equation}
 Solving (\ref{ep}) yields, 
 $$\epsilon^{c}=\dfrac{1}{48\pi}(2u^{\alpha}\nabla^{\beta}~-~u^{\beta}\nabla^{\alpha})\nabla_{\alpha}u_{\beta}+ \bar{C}T^{2}$$
 \begin{equation}
 \label{p}
 p^{c}~=~ -\dfrac{1}{48\pi}u^{\alpha}\nabla^{\beta}\nabla_{\alpha}u_{\beta}
 \end{equation}
 
 We reiterate that $\epsilon^{c}$ and $p^{c}$ are introduced to mimic the actual energy density $\epsilon$ and pressure $p$ so that the constitutive relation for the anomalous sector bears a resemblence to the usual ideal structure.
 
 Let us wind up by stating that we have  developed a novel way of analyzing chiral hydrodynamics where special properties of chirality in two dimensions were exploited. Apart from other features discussed above, this explains the functional structure of the velocity factor in (\ref{tmunu}, \ref{CT}, \ref{mufin}). It is the unique form that ensures the chirality of $T_{\mu\nu}$. This was elaborated by explicitly working out the constitutive relation for an ideal(chiral) fluid. Given in(\ref{mufin}) this is a new structure  that is different from the familiar relation(\ref{consti}) for a nonchiral ideal fluid. 
\section{Conclusion}

Gravitational anomalies in 1+1 dimensions have played a ubiquitous role in physics. Besides producing surprising connections between physics and mathematics, they have been  used extensively to understand the Hawking effect in black holes\cite{kulkarni, 2008 ,iso, rbanerjee, fulling} or the thermal Hall effect in topological insulators\cite{stone}. Very recently their influence in hydrodynamics has generated considerable interest\cite{arxiv:1203, jain, superfluid, wesszumino, valle, cones}. The two dimensional example, in particular, is somewhat unique, because of its special features.

In this paper we have developed a new formulation for two dimensional chiral hydrodynamics with anomalies. The special  characteristics of both chirality and two dimensions have been taken  into account. While the specific nature of two dimensions ensures that the formulation yields exact results, chirality puts additional restrictions on the structure of the constitutive relations. These restrictions are algebraic in nature and are thus valid even for an ideal(chiral) fluid. We have explicitly given the form for the constitutive relation(see \ref{mufin}) for an ideal(chiral) fluid. This new structure helps to understand the eventual nature of the constitutive relation for quantum chiral hydrodynamics. At the quantum level, chirality implies the presence of both diffeomorphism and conformal(trace) anomalies. It was reassuring to note that the constitutive relation (\ref{CT}) for quantum chiral hydrodynamics was compatible with both types of anomalies as well as the algebraic restriction imposed by chirality in two dimensions. Finally, the constitutive relation involves an arbitrary constant manifesting a one parameter class of solutions. A particular value of this constant ($\bar{c}=\dfrac{\pi}{12}$) exactly reproduces the results found by the gradient expansion approach while this consistency check definite supports our analysis, we feel that the explanation of the functional form of the velocity factors in (\ref{mufin})in terms of chirality is an important point that was missing in the existing literature.

 This analysis illuminates a similarity between hydrodynamics and quantum field theory. Two dimensional models in QFT are exactly solvable and using the technique of bosonisation, may be put in a form that resembles the free theory but with a different normalisation of the parameters\cite{abdalla}. Something analogous happens here also. The exact constitutive relation for two dimensional anomalous chiral hydrodynamics is put in the corresponding form of an ideal chiral fluid but with a different normalisation  of the parameters.
 
 It is possible to extend or generalise the present work. For example, one can consider the case of a charged chiral fluid so that the analysis has to be performed in the presence of gravitational as well as gauge anomalies. Another aspect would be to drop the chirality condition  and treat a nonchiral  fluid in the presence of anomalies. This case is fundamentally different from chiral hydrodynamics considered here because the lack of unidirectional property ensures that both diffeomorphism and coformal anomalies do not coexist. There would be a diffeomorphism anomaly but no trace anomaly or vice-versa. Since for gravitational theories diffeomorphism invariance is more fundamental it is retained at the expense of a conformal anomaly. Its consequences for hydrodynamics have been considered, albeit briefly, in\cite{cones,oz} within the derivative expansion scheme. Nevertheless, here also one expects exact results in the two dimensional case. Implications of this analysis to higher dimensions is yet  another possibility to be explored. Work in these directions is in progress and will soon be reported\cite{r2013}.


\end{document}